\documentclass[prl,twocolumn,superscriptaddress,showpacs,nofootinbib]{revtex4}
\usepackage{epsfig}

\newcommand{\be}{\begin{equation}}
\newcommand{\ee}{\end{equation}}
\renewcommand{\r}{{\bf r}}

\newcommand{\q}{{\bf q}}

\newcommand{\ep}{\epsilon}

\begin{document}


\title{Non exponential quasiparticle decay  and  phase relaxation

 in low
dimensional conductors }
\author{G. Montambaux}
\affiliation{Laboratoire de Physique des Solides, CNRS UMR 8502,
  Universit\'e Paris-Sud, 91405 Orsay, France}
\author{E. Akkermans}
\affiliation{Department of Physics, Technion Israel Institute of Technology,
  32000 Haifa, Israel}


\begin{abstract}
 We show that  in low dimensional disordered
conductors, the quasiparticle decay and the relaxation of the
phase are not exponential processes. In the quasi-one dimensional
case, both behave at small time  as $e^{- (t/\tau_{in})^{3/2}}$
where the inelastic  time $\tau_{in}$, identical for both
processes, is a power $T^{2/3}$ of the temperature. The non
exponential quasiparticle decay results from a modified derivation
of the Fermi golden rule. This result implies the existence of an
unusual  distribution of relaxation times.

\end{abstract}


\date{\today}

\maketitle


The issue of dephasing in the presence of electron-electron
interactions in disordered conductors is of great importance in
mesoscopic physics. This problem, first addressed by Altshuler,
Aronov and Khmelnitskii (AAK) \cite{Altshuler8213}, has been
recently revisited in the light of a new set of experiments
\cite{Mohanty9713,Gougam99,Esteve02}. A related problem is the
understanding of the time evolution of a quasiparticle state due
to electron-electron interaction, which governs the relaxation
towards thermal equilibrium \cite{Altshuler8513}. Since coherent
effects in disordered systems result from   the coherent pairing
of two scattering amplitudes defined for a given quasiparticle
state, the coherence is lost once this state has relaxed. Thus it
seems natural to assume that quasiparticle and phase relaxations
are of the same nature \cite{Imry} and are driven by the same time
scale. In this letter, we show that in low dimensional disordered
conductors and in particular for quasi-one-dimensional
(quasi-$1d$) wires, both relaxations are faster than exponential
and are driven by the same characteristic time \cite{am}. This non
exponential behavior reflects the existence  of a distribution of
relaxation times.  Such a non exponential decay is unusual in quantum condensed matter physics but more frequent in the context of molecular relaxation processes and in glassy systems. Stretched and compressed exponentials are mostly used as a way to fit unusual relaxations but no microscopic basis can be assigned to account for this behavior \cite{phillips}. Here, we derive it from a new treatment of the Fermi golden rule.

We shall first consider the quasiparticle decay, using the Fermi
golden rule \cite{Abrahams} . We show that, due to screened
Coulomb interactions with small energy transfer, the relaxation
rate is not constant, implying a non exponential decay. This
results from a key step in the Fermi golden rule which stems that
the transitions conserve energy within $\hbar /t$ where $t$ is the
duration of the perturbation.
Usually this constraint is
of no practical importance and   energy conservation is described by a delta
function. Here we show that this is no longer possible.
As a result, we find that  for quasi-1$d$\,  wires, the
probability for
 a quasiparticle to stay in its initial state behaves, at small times,  as ${\cal
P}(t) = e^{- \beta (t/\tau_{in})^{3/2}}$.  The temperature
dependence of the inelastic time is $\tau_{in}(T) \propto
T^{-2/3}$ and $\beta$ is a numerical constant.

Then, we shall come to the relaxation of phase coherence. It is
given by the average $\langle e^{i \Phi(t)} \rangle$ of the
relative phase $\Phi(t)$ between time reversed trajectories.
Starting from the AAK calculation, we   show that the phase
relaxation is also non-exponential and that, at small times, it
behaves like $\langle e^{i \Phi(t)} \rangle \simeq e^{-
(t/\tau_{\phi})^{3/2}}$ where the phase coherence time $\tau_\phi$
is proportional  to the quasiparticle decay
time $\tau_{in}$.

 We start by considering the decay of a quasiparticle, recalling first  some
 known  features of the derivation.
Using the Fermi golden rule, the quasiparticle  lifetime can be
written in terms of a kernel $K(\omega)$ which is the  average
over disorder of the  squared matrix element of the screened
Coulomb interaction \cite{am,eiler}:
\be {\hbar \over \tau_{in}(T)}= 8 \pi \nu_0^3 \int_0^\infty
d\omega K(\omega){\omega \over \sinh \beta \hbar \omega} \ \ ,
\label{tauinkernelT}\ee
 where $\nu_0$ is the total density
of states per spin direction. The temperature dependence results from
the  occupancy of the initial and final quasiparticle states. Upon disorder averaging, the kernel $K(\omega)$
is obtained as the squared  product  of the dynamically screened
interaction and of a long range contribution called diffuson. As a result, we have
\cite{Altshuler8513}~:
\be K(\omega)= {1 \over 4 \pi^2 \nu_0^4} \sum_\q {1 \over \omega^2
+D^2
q^4} \ \ . \label{Ksum} \ee %
 The diffusive nature of the
electronic motion  implies  a strong dependence of the transition
probability upon  the space dimensionality $d$ that appears in the
sum over the modes $\q$. The kernel $K(\omega)$ then depends on
the space dimensionality $d$ and it is given by
\cite{Altshuler8513} $ K(\omega)= (\alpha_d / 16 \nu_0^4 \omega^2
) \left( \omega / E_c \right)^{d/2} $ where $\alpha_1=
\sqrt{2}/\pi^2$, $\alpha_2= 1/2\pi^2$, $\alpha_3=\sqrt{2}/2\pi^3 $
and $E_c=  D/L^2$ is the Thouless frequency. For $d=3$, the
integral  in (\ref{tauinkernelT})  is convergent so that $\tau_{in}(T)$
is well defined  and behaves like $T^{-3/2}$. However for $d \leq
2$, the integral in (\ref{tauinkernelT}) diverges at low energy
transfer $\omega$. To cure this divergence, it is commonly argued
that the low frequency cut-off needed is $1/ \tau_{in}$ itself, since no energy transfer can be smaller
than $\hbar / \tau_{in}$. Consequently, the lifetime is solution
of a self-consistent equation whose solution in $d=1$ is~:
\be {1 \over \tau_{in}(T)} = \left({ T e^{2} {\sqrt D}\over S
\sigma \hbar^2 }\right)^{2/3} \ \ . \label{tauphi1D4}\ee
where the conductivity is $\sigma= 2 e^2 \nu_0 D / (LS)$. This
temperature dependence has been first obtained by Altshuler and
Aronov \cite{Altshuler8513}.

We  argue here  that this divergence is indeed the
signature of a {\it new behavior for the quasiparticle decay}. We
prove that this decay is actually {\it non exponential}. The
crucial point in our argument is that it is not correct to cut-off
the integral at $1/\tau_{in}$. To grasp the relevance of this
statement, it is important  to recall the Fermi
golden rule prescription
 namely that,  after a time $t$, the range of accessible
states is limited to energies larger than $\hbar / t$, {\it not}
$\hbar /\tau_{in}$. This leads to the replacement of
Eq.(\ref{tauinkernelT}) by the following expression for the
disorder averaged transition probability ${\cal P}^{(2)} (t)$
towards final states, calculated  up to  second-order in
perturbation~:
\be  {\cal P}^{(2)} (t)  = {\pi \alpha_d T \over 2 \nu_0 \hbar^2
} \ t  \int_{1/t}^{T / \hbar} {d \omega \over \omega^2} \left(
{\omega \over E_c}\right)^{d/2} \label{tauphi1D2}\ee
where, for simplicity, the thermal factor  has been replaced by an
upper cut-off at  $\hbar \omega \sim T$. We have used the above
expressions for the kernel in $d$ dimensions.
In one  dimension,  this leads immediately
to a $t^{3/2}$ power law~:
$   {\cal P}^{(2)} (t)
= { \sqrt{2}  T \over \pi \nu_0 \hbar^2 \sqrt{E_c}} t^{3/2}
\label{tauphi1D3} $
so that the quasiparticle relaxation  is not  exponential.

Let us prove now this qualitative statement, coming back to the
derivation of the Fermi golden rule. A given initial quasiparticle state
$\alpha$ interacts  with a quasiparticle of energy $\ep_\gamma$,
leading to two quasiparticles  of  final energies $\ep_\beta$ and
$\ep_\delta$. As known from quantum mechanics textbooks
\cite{messiah}, the transition probability towards final states
is~:
\be {\cal P}^{(2)}_\alpha(t)=2 \sum_{\beta\gamma\delta} |
U_{\alpha \gamma, \beta \delta}|^2f_t (
{\ep_\alpha+\ep_\gamma-\ep_\beta-\ep_\delta \over \hbar})
\label{Wa11}\ee
 where $U_{\alpha
\gamma, \beta \delta}$ is the matrix element of the interaction.
The function $f_t(\Delta \omega)$ of width $\pi/t$ is given by
\cite{messiah}
\be f_t(\Delta \omega)=\left(\sin \Delta \omega t/2 \over \Delta
\omega/2 \right)^2 \ \ . \label{ftDw} \ee
Its maximum is  equal to $t^2$ and its integral is $2 \pi t$.
Usually,  this function can be approximated  by  $2 \pi t\
\delta(\Delta \omega)$, so that the decay is linear in $t$, and
the prefactor is the inverse quasiparticle time.

The main idea here is that this approximation is not always valid.
 To see this, we first calculate the disorder average of Eq. (5)
using standard methods \cite{am}.  The new input is that the
energy of the initial and final states may differ by a small
amount $\Delta \omega$ of order $1/t$, as explicitly seen on
Eqs.(\ref{Wa11}) and (\ref{ftDw}). As a result, the diffusons and
dynamically screened interactions which enter in the kernel $K$
have to be taken at different  frequencies $\omega_\pm= \omega \pm
\Delta \omega/2$. This immediately leads to a  kernel $K_{\Delta
\omega}(\omega)$ which now depends on the energy difference
$\Delta \omega$~:
\be K_{\Delta \omega}(\omega)={1 \over 4 \pi^2 \nu_0^4 }\sum_{\q
\neq 0} {1 \over \sqrt{(\omega_+^2 + D^2 q^4)(\omega_-^2 +D^2
q^4)}}  \label{W2q2}
\end{equation}
instead of (\ref{Ksum}). This kernel yields to a transition  probability of the form~:
\begin{equation}
 {\cal P}^{(2)}(t)={4 \nu_0^3 T  \over \hbar^2}
\int_{0}^{T/ \hbar} d \omega  g_t(\omega) \ \ , \label{PdetwDw}
  \ee
\begin{equation}
\mbox{where} \qquad g_t(\omega)=  \int_{-\infty}^\infty  d
\Delta\omega\ f_t(\Delta \omega) K_{\Delta \omega}(\omega)
  \ \ .   \label{PdetwDw3}
  \ee
Letting  $f_t(\Delta \omega)= 2 \pi t \delta(\Delta \omega)$ leads
immediately  to  the usual result, namely a behavior of $ {\cal
P}^{(2)}(t)$  linear in $t$ and defines $1/\tau_{in}$ provided the
integral is convergent.

However in $1d$, when  $\Delta \omega=0$, $K_{\Delta
\omega=0}(\omega) = K(\omega) \propto 1/\omega^{3/2}$  and the
integral on $\omega$
  becomes divergent.
It is thus crucial to keep the  full expression  of $f_t(\Delta
\omega)$. Doing this, we find that for  $\omega \gg 1/t$,  we
still have $g_t(\omega)= 2 \pi t K(\omega) \propto
t/\omega^{3/2}$,  while  for $\omega \ll 1/t$, it is easy to check, since $f_t(0)=t^2$, that the function $g_t(\omega)\propto t^2
/\sqrt{\omega}$, so that its integral near zero frequency is
indeed convergent. More precisely,  we can show that $g_t(\omega)$
is of the form $g_t(\omega)=2 \pi t K(\omega) h(\omega t)$ where
the function $h(\omega t)$ is linear for small argument and tends
to $1$ for large argument.  This function is calculated
numerically and is shown on Fig. \ref{coupure}. The
quasiparticle decay probability  takes the form
\begin{equation}
 {\cal P}^{(2)}(t)={  T  t \over \pi \sqrt{2}\hbar^2 \nu_0 \sqrt{E_c}}
\int_{0}^\infty   {d \omega \over \omega^{3/2}} h(\omega t)\ .
   \label{PdetwDw2}
  \ee
\begin{figure}
\centerline{ \epsfxsize 6cm \epsffile{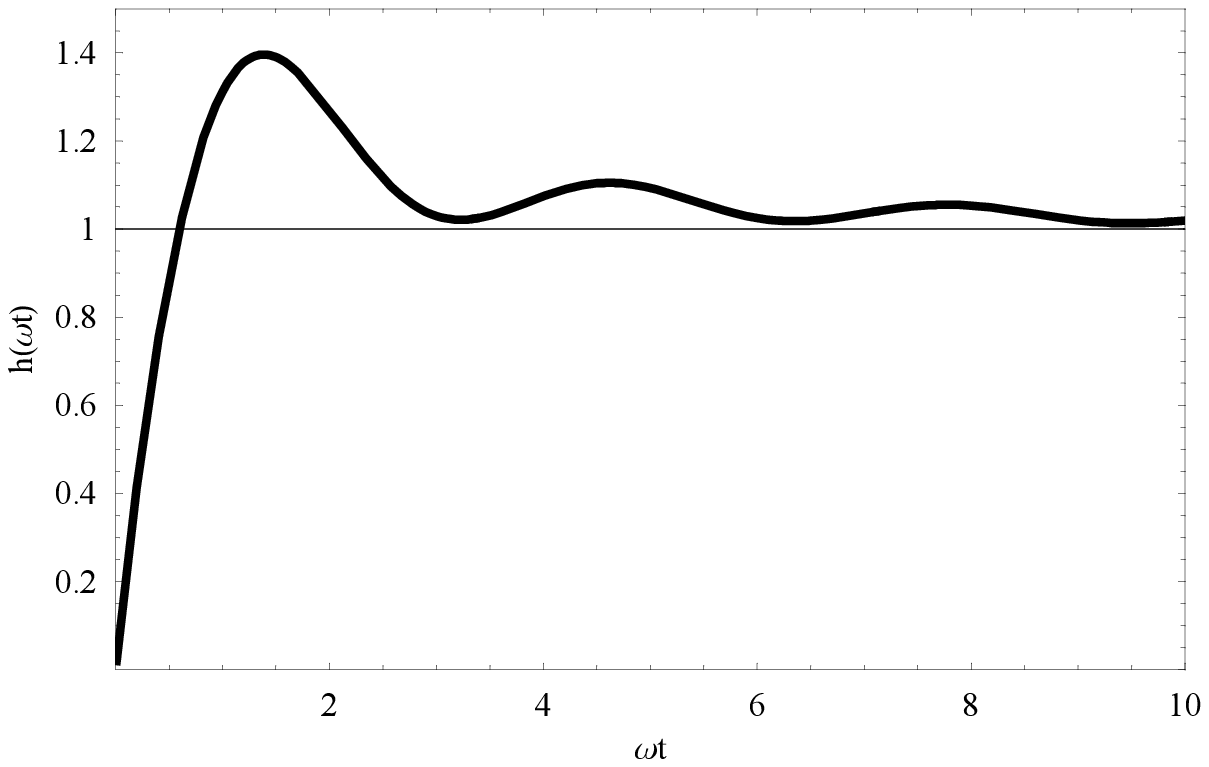} } \caption{\it
Plot of the function $h(\omega t)$. It vanishes for small
argument, justifying the cut-off of order $1/t$ in Eq.
(\ref{tauphi1D2}).} \label{coupure}
\end{figure}
We have thus proven that the cut-off in (\ref{tauphi1D2}) appears as a natural consequence of  a proper use of the Fermi golden
rule. More precisely, because of the function $h$ which naturally
 provides  a lower cut-off of order $1/t$, the integral now
converges and
\be {\cal P}^{(2)}(t) ={\sqrt{2} e^2 \sqrt{D} T  \over \pi \sigma
S \hbar^2 } \ t^{3/2} \int_0^\infty {d x \over x^{3/2}} h(x) \ .
\ee
 At small times, the survival
probability ${\cal P}(t)$,  {\it i.e.} the probability that a
quasiparticle stays in its original state, is given by ${\cal
P}(t)=1-{\cal P}^{(2)}(t)$.  By exponentiating this relation,
we obtain :
\be {\cal P}(t) = e^{- \beta \left({t \over
\tau_{in}}\right)^{3/2}}  \ . \label{prelax} \ee
 The survival probability is thus given by a  compressed
exponential characterized by the inelastic time
Eq.(\ref{tauphi1D4}) and where $\beta \simeq 5.83 \sqrt{2}/ \pi$.
Similarly, in two dimensions for a film of thickness $a$, using  (\ref{tauphi1D2}), one finds a logarithmic correction $  {\cal P}(t) \sim e^{-{t \over
\tau_{in}}{ 1 \over  \ln T t}} $ with $\tau_{in}^{-1} (T) \propto
{ e^{2} T \over \hbar^2 \sigma a}$. These temporal behaviors
constitute one of the main results of this paper.
 We emphasize again that there  are direct consequences of  the
Fermi golden rule prescription according to which the energy is  conserved,
{\it not within the decay rate $\hbar /\tau_{in}$, but
rather within the inverse time $\hbar /t$}
 \cite{messiah}.

Now, one can wonder whether this peculiar behavior of the energy
relaxation has its signature in the time dependence of the phase
relaxation of coherent effects in weakly disordered systems. These
effects result from  the coherent pairing of two scattering
amplitudes defined for a given quasiparticle state. In particular,
we consider pairs of time reversed trajectories (the cooperon) as
it appears in the weak-localization correction to the
conductivity. The relaxation of the cooperon is driven by a
characteristic time $\tau_{\phi}$ called the phase coherence time.
It seems quite intuitive that, as far as Coulomb interactions are
involved, the quasiparticle decay  and the cooperon relaxation are
related. Thus the question arises of the relation between the
inelastic time $\tau_{in}$ and the phase coherence time
$\tau_{\phi}$. We shall now show that these two relaxation
processes are indeed identical and characterized by the same time
scale.

To that purpose, we consider the time relaxation of the cooperon
  by replacing the Coulomb interaction by a classical fluctuating
potential $V(\r,\tau)$ whose characteristics are determined by the
fluctuation-dissipation theorem \cite{Altshuler8213}. The cooperon
contribution to the return probability can be written under the
form~:
\be  P_c(\r,\r,t)=   P_c^{(0)}(\r,\r,t) \left\langle e^{i
\Phi(\r,t) }\right\rangle_{T,{\cal C}} \label{Pcfluc1}  \ \ \  \ee
where $P_c^{(0)}$ is the cooperon  in the absence of the
fluctuating  potential and  $\Phi=\Phi(\r,t)$ is the relative
phase of a pair of time reversed trajectories at time $t$~:
\be \Phi ={1 \over \hbar} \int_{0}^{t}
[V(\r(\tau),\tau)-V(\r(\tau),\overline{\tau})] d\tau
\label{PhiVV}\ee
This expression is valid in the eikonal approximation {\it i.e.}
for a slowly varying potential whose effect is to multiply the
disorder averaged Green function by a phase term proportional to
the circulation of $V(\r(\tau), \tau)$ along the trajectory
between the times $0$ and $t$.  We
 define  $\overline{\tau}=t-\tau$.

 The dephasing $\Phi$ is accumulated along the diffusive
electronic trajectories paired in the cooperon. One of them
propagates in the time interval $0 \leq \tau \leq t$ whereas its
time reversed counterpart propagates from $\tau=t$ to $\tau=0$. We
denote by $\langle \cdots \rangle_{T,{\cal C}}$ the average taken
both over the distribution of the diffusive trajectories ($\langle
\cdots \rangle_{{\cal C}}$)
 and over the thermal fluctuations ($\langle \cdots \rangle_{T}$) of the electric potential.
 The latter are Gaussian so that the thermal average
   $\left\langle e^{i \Phi }\right\rangle _{T} = e^{- {1 \over 2}
\left\langle\Phi^2 \right\rangle_T}$. Using (\ref{PhiVV}) and the
fluctuation-dissipation theorem in the classical limit ($\beta
\hbar \omega \ll 1$), namely, $ \left\langle V(\q, \omega) V(-\q,
\omega) \right\rangle_{T} = {2 e^{2} T \over \sigma q^{2}}$,   we
obtain for the dephasing the following expression
\begin{equation}
   \left\langle \Phi^2 \right\rangle_T={4 e^2 T \over
 \sigma \hbar^2} \int_{0}^{t}d\tau
 \int {d\q \over (2 \pi)^d} {1 - \cos \q.
 (\r(\tau)-\r(\overline{\tau})) \over  q^{2}}
\label{PhiVV3}\ee
The average $\langle \cos \q.
 (\r(\tau)-\r(\overline{\tau})) \rangle$ over the diffusive trajectories
  of time $t$ is $e^{- 2 D q^2 \tau|1 - 2 \tau/t|}$. For a quasi-1d wire, the integrations
  over $q$ and $\tau$ lead to
 \be \left\langle \Phi^2
\right\rangle_{T,{\cal C}}=  {\sqrt{\pi}  e^2 T \over 2  \hbar^2
\sigma S }\sqrt{ D}\  t^{3/2}  = {\sqrt{\pi} \over 2} \left({t
\over \tau_{in}}\right)^{3/2}\label{Phi2}\ee Assuming first that
$\left\langle e^{-{1 \over 2}\left\langle \Phi^2 \right\rangle_T}
\right\rangle_{\cal C} \simeq e^{-{1 \over 2}\left\langle \Phi^2
\right\rangle_{T,{\cal C}}}$, we obtain for the cooperon, at small
time, the compressed exponential behavior
 \be
\left\langle e^{i \Phi }\right\rangle_{T,{\cal C}} \simeq e^{-{
\sqrt{\pi}/4}
 \left({t / \tau_{in}}\right)^{3/2}} \label{gauss}\ee
identical to  the energy relaxation (\ref{prelax}) and with the
same characteristic time $\tau_{in}$ given by (\ref{tauphi1D4}). A
similar behavior for the phase relaxation has been also found in
\cite{Stern13}.
It is interesting at this stage to compare (\ref{PhiVV3}) with
(\ref{tauphi1D2}) obtained for the transition probability of a
quasiparticle state. Although these expressions behave similarly,
the convergence in Eq.(\ref{tauphi1D2}) results from a cut-off at
small $\omega$ whose origin is in the Fermi golden rule
prescription namely, that among the large number of accessible
states in $d=1$, only those with  energy transfer larger than
$\hbar /t$ are accessible after a time $t$. This low energy cutoff
does not exist in (\ref{PhiVV3}) and  the convergence results from
the compensation between the two terms in the bracket that
describe respectively the contributions of the correlations
$\langle V(\r(\tau), \tau) V(\r(\tau'), \tau') \rangle_{T} $ and
$\langle V(\r(\tau), \tau) V(\r(\tau'), {\overline \tau}')
\rangle_{T} $ to the cooperon.

The result (\ref{gauss}) is not fully correct since we have
approximated the average $\langle \cdots\rangle_{\cal C}$ of the
exponential by the exponential of the average. Using the
functional integral approach presented in \cite{Altshuler8213},
there is a way to derive an expression for the phase relaxation
valid at all times by considering the Laplace transform $
P_\gamma(r,r)=\int dt P_c(r,r,t) e^{-\gamma t} $ of the cooperon.
In quasi-$1d$, one has :
\be P_\gamma(r,r)
=-{1 \over 2} \sqrt{\tau_{in}
\over D} {\mbox{Ai}(\tau_{in}/\tau_\gamma) \over
\mbox{Ai}'(\tau_{in}/\tau_\gamma)}\label{PgC2g} \ee
where  $\mbox{Ai}$ et $\mbox{Ai}'$ are respectively the Airy
function and its derivative  \cite{Abramovitz13} and
$\tau_\gamma=1/\gamma$. The probability $P_c(r,r,t)$ in
(\ref{Pcfluc1}) can thus be obtained from the inverse Laplace
transform of (\ref{PgC2g}). Since,  in the quasi one-dimensional
limit, one has $P_c^{(0)}(r,r,t)=1/\sqrt{4 \pi D t}$, the
dephasing term $\left\langle e^{i \Phi} \right\rangle_{T,{\cal
C}}$ is a function of $t / \tau_{in}$ that satisfies
\be   \int_0^\infty  {dt \over \sqrt{t}} \left\langle e^{i
\Phi} \right\rangle_{T,{\cal C}} e^{-t/\tau_\gamma}=-\sqrt{\pi \tau_{in}}
{\mbox{Ai}(\tau_{in}/\tau_\gamma) \over
\mbox{Ai}'(\tau_{in}/\tau_\gamma)}\label{airy4}\ee
The inverse Laplace transform is obtained by noticing that the
Airy function and its derivative are analytic and non meromorphic
functions whose zeroes lie on the negative real axis. Then, by
performing the integral in the complex plane with the residues
$\mbox{Res}(e^{st} \mbox{Ai}(s) / \mbox{Ai}'(s)) = e^{-|u_{n}|t} /
|u_{n}|$ where the $u_n$ are the zeros of $\mbox{Ai}'(s)$ given at
a very good approximation by $|u_{n}| = \left( {3 \pi \over 2} (n
- {3 \over4} )\right)^{2/3} $ \cite{Abramovitz13}, we obtain the
analytic function
\be \left\langle e^{i \Phi} \right\rangle_{T,{\cal C}}= \sqrt{{\pi
t \over \tau_{in}}} \sum_{n=1}^\infty {e^{- |u_n| t/\tau_{in}}
\over |u_n|} \label{fun} \ee
 At small times $t < \tau_{in}$,  it   behaves like
Eq.(\ref{gauss}).  At large time, the relaxation is driven by the
first zero of the $\mbox{Ai}'$ function, namely $ \left\langle
e^{i \Phi} \right\rangle_{T,{\cal C}} \simeq \sqrt{\pi t /
\tau_{in}}
 e^{-|u_1| t/\tau_{in}} / |u_1| $ with $|u_1| \simeq 1.019$.
Clearly, the relaxation (\ref{fun}) is never exponential. It
appears as a  distribution of relaxation times
$\tau_{in} / |u_{n}|$ which is at the origin of the rather unexpected compressed
exponential behavior of the quasiparticle decay and of the cooperon phase relaxation. The expression (\ref{fun}) constitutes one of the main results of this paper.
\begin{figure}
\centerline{ \epsfxsize 7cm \epsffile{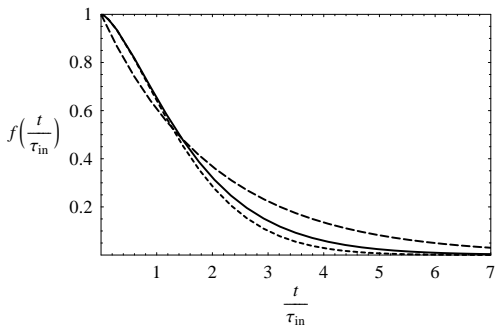} }
\caption{\it Behaviour of $ \langle e^{i \Phi(t)}\rangle_{T,{\cal
C}}$. The continuous line is the exact result (\ref{fun}). The
dotted line is obtained from the small time expansion
(\ref{gauss}). The dashed line shows the exponential fit $e^{-  t
/ 2\tau_{in}}$.} \label{airyplot}
\end{figure}
The question arises of how this
behavior could show up experimentally. It has been stressed by
Pierre {\it et al.} \cite{Esteve02}, that the
Laplace transform (\ref{PgC2g}) is well
approximated
 by the relation $ P_\gamma(r,r)= (1 / 2\sqrt{D}) \left( {1 \over
\tau_\gamma}+ {a \over  \tau_{in}} \right)^{1/2}$ where $a \simeq
1/2$ is an adjustable numerical constant. This approximation
corresponds to an exponential relaxation $\langle e^{i
\Phi(t)}\rangle_{T,{\cal C}} \simeq e^{-  t/ 2\tau_{in}}$ that is
clearly at odd with the behavior (\ref{fun}) (see   Fig.
\ref{airyplot}). However,  the difference between the exact
relation and the exponential approximation is difficult to see
experimentally. The time $\tau_\gamma$ accounts for other
 processes such as
 the decay rate in a magnetic field which, for a wire of section $S$,
 is given by $\tau_\gamma^{-1}=\tau_B^{-1}=
D S^2 e^2 B^2/3 \hbar^2$ \cite{AA}. A possibility to probe the $t^{3/2}$ behavior at small time is to study the limit $\tau_B \ll \tau_{in}$ where the weak localization correction to the conductivity is
 \be \Delta \sigma =  -  {e^2 \over \pi \hbar} \sqrt{D \tau_B}
\left[ 1 - {1 \over 4} \left({\tau_B \over
\tau_{in}}\right)^{3/2}\right] \label{sigC1} \ee
The power-law dependence of the correction term in (\ref{sigC1})
is a direct signature of the  $t^{3/2}$ behavior of the relaxation
at small time. The asymptotic behavior $ \Delta \sigma (B) -
\Delta \sigma(B \rightarrow \infty)
 \propto T / B^4 $ reflects both the
$T^{-2/3}$ dependence of $\tau_{in}$ and the non exponential
$t^{3/2}$ phase relaxation. Another direct probe of the non exponential relaxation of the
phase should may also be provided by the behavior of the {\it ac}
conductivity $\sigma (\omega)$.



This research was supported in part by the Israel Academy of
Sciences, by the Fund for Promotion of Research at the
Technion, and by the French-Israeli Arc-en-ciel program.

    \end{document}